\documentclass[prb,aps,twocolumn,showpacs,preprintnumbers,superscriptaddress,amsmath,amssymb]{revtex4-1}
\usepackage{graphicx}% Include figure files
\usepackage{dcolumn}% Align table columns on decimal point
\usepackage{bm}% bold math
\usepackage{color}
\begin{document}

\title{Site and orbital selective correlations in $\beta$-Pu}

\author{W. H. Brito}
\affiliation{Department of Physics and Astronomy, Rutgers University, Piscataway, New Jersey 08854, USA.}
\author{G. Kotliar}
\affiliation{Department of Physics and Astronomy, Rutgers University, Piscataway, New Jersey 08854, USA.}
\affiliation{Condensed Matter Physics and Materials Science Department, Brookhaven National Laboratory, Upton, New York 11973, USA.}

\begin{abstract}
 We investigate the electronic structure of the highly anisotropic $\beta$ phase of metallic plutonium, within the combination of density functional theory (DFT) and dynamical mean field theory (DMFT). Its  crystal structure gives rise to  site  and orbital selective  electronic correlations, with coherent Pu-5$f_{5/2}$ states and very incoherent Pu-5$f_{7/2}$ states. The Hund's coupling is essential for determining the level of correlations of electrons in Pu-5$f$ states, and for the quasiparticle multiplets features in the Pu-5$f$ spectral function.  
\end{abstract}

\maketitle

\section{Introduction}

Pure plutonium (Pu) exhibits the richest phase diagram among the simple metals with six allotropic phases in a relatively narrow temperature range (between 100 K and 756 K).~\cite{moore_review} Most of the theoretical focus has been on the $\delta$ phase, which  exhibits clear correlation effects, as for example a very enhanced specific heat~\cite{lashleyPRL,havelaPRB} and a  negative thermal-expansion coefficient, without localized static magnetic moments.~\cite{lashley,heffnerPRB} Furthermore, early density functional theory (DFT) calculations failed to predict the equilibrium volume of elemental Pu and its paramagnetic phase.~\cite{dftpu1,dftpu2} On the other hand, dynamical mean field theory (DMFT) studies have been very successful in understanding the properties of elemental Pu,~\cite{elihu,xidai_Science,marianetti_PRL,amadon_PRB} and provided a physical picture of its $\delta$ phase as a mixed valence metal,~\cite{shimOCAPu} which has been recently confirmed experimentally.~\cite{janoschek}

There are very few DMFT studies of the other phases of Pu.  Only recently, the  ground state, i.e. the $\alpha$ phase of elemental Pu, which has  a monoclinic crystal structure with eight  distinct crystallographic sites, was studied using the combination of DFT with DMFT.~\cite{willsNatCom} Previous DFT calculations pointed out that this highly complex structure occurs in part due to a Peierls distortion associated with the Pu-5$f$ band.~\cite{soderling_PRBPeierls} Zhu et al.~\cite{willsNatCom} revealed that $\alpha$ phase, which is 25$\%$ more dense than the $\delta$ phase, is not weakly correlated, and in fact exhibits site dependent correlations, with some sites having large Kondo scales while others have large mass renormalizations. 
For a recent review see.~\cite{review_soderlind} Theoretical overviews of all the phases of Pu are available using slave boson methods,~\cite{lanataPRX} density functional methods,~\cite{soderlindPRL} and from thermodynamic data.~\cite{wallacePRB} But, up to now, no detailed investigation of the electronic structure of phases other than $\alpha$ and $\delta$ is available beyond density functional theory. 

In this work we study the electronic structure of the $\beta$ phase of metallic Pu. The $\alpha$ to $\beta$ transition, is the first phase transition that Pu undergoes upon raising temperature accompanied by a ten percent volume expansion. The $\beta$ phase is monoclinic and has seven inequivalent sites and unusual elastic properties,~\cite{suzukiPRB} but there are very few experimental reports on its electronic properties,~\cite{havela_puos} and the degree of localization of the Pu-$5f$ electrons in this phase is not known. 

An important question is the physics of orbital differentiation, which in the extreme case can lead to an orbital selective Mott transition.~\cite{anisimov_ruth} This has been extensively studied in materials with $d$ electrons,~\cite{ziping,lanata_iron,demedici_PRL,miao} but is only recently beginning to be explored for $f$ electron systems. More recently, orbital differentiation have been observed  in UO$_2$,~\cite{lanata_uo2,huang_uo2} in californium (Cf),~\cite{huang_cf} and in PuCoGa$_5$.~\cite{whbpu115} 
Furthermore, $\beta$-Pu is an ideal playground for exploring how the strength of the correlations vary with different sites, and between different orbitals in the same site.
According to our calculations there are some site selective electronic correlations in $\beta$-Pu, with coherent Pu-$5f_{5/2}$ states and very incoherent Pu-5$f_{7/2}$ orbitals. We also find that the Hund's rule coupling is essential for determining the level of correlations in this material, and to describe the quasiparticle multiplets in the Pu-5$f$ spectral function, which are fingerprints of the physics of Hund's-Racah metals.~\cite{shick_racah, Lichtenstein_Racah}

\section{Computational Methods}
\label{method}

Our calculations were performed using the fully charge self-consistent DFT+embedded-DMFT approach,~\cite{hauleWK} as implemented in EDMFTF code.~\cite{haulepage}
The DFT calculations were performed within Perdew-Burke-Ernzehof generalized gradient approximation (PBE-GGA),~\cite{pbe} as implemented in Wien2K package.~\cite{wien} To solve the DMFT effective impurity problem we used the vertex-corrected one-crossing approximation (OCA)~\cite{pruschke} with the on-site Coulomb repulsion $U = 4.5 $ eV and Hund's coupling $J = 0.512$ eV. We emphasize that these values of $U$ and $J$ were used successfully to describe the valence-fluctuating ground state of $\delta$-Pu within our implementation,~\cite{janoschek} which in turn takes into account all the itinerant and correlated states within a 20 eV energy window around the chemical potential.
In addition, we used the standard fully localized-limit form~\cite{anisimovEdc} for the double-counting correction term, with $n_{f}^{0}=5.2$, which is the average occupation of Pu-5$f$ states in the $\delta$-Pu.~\cite{shimOCAPu}

\section{Results and Discussions}

\subsection{Site-selective correlations}

\begin{figure}[!htb]
\includegraphics[scale=0.45]{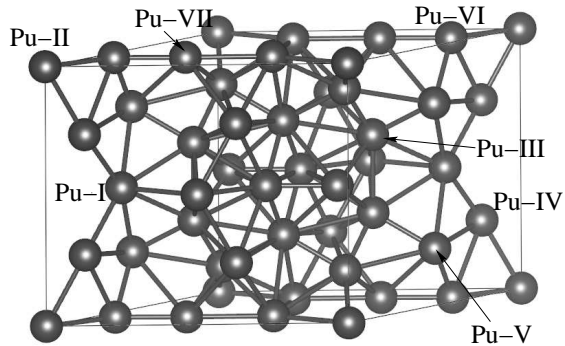}
 \caption{Crystal structure of $\beta$-Pu (space group $I2/m$), where the Pu atoms are represented as black spheres. The seven inequivalent Pu sites are schematically indicated by the arrows.}
\label{fig:crystal_bpu}
\end{figure}

At ambient pressure and temperatures below 488~K, elemental Pu exhibits two low-symmetry monoclinic phases, namely the $\alpha$ and $\beta$ phases. The $\alpha$ phase is stable up to 398 K and crystallizes with a monoclinic unit cell of 16 atoms and eight distinct Pu crystallographic sites. The $\beta$ phase has a monoclinic body-centered structure with 34 atoms within the unit cell and seven distinct Pu sites.~\cite{ellinger} In Fig.~\ref{fig:crystal_bpu} we illustrate the experimental $\beta$-Pu crystal structure which was used in our calculations. The inequivalent Pu sites are schematically depicted in the same figure. 

\begin{figure}[!htb]
\includegraphics[scale=0.42]{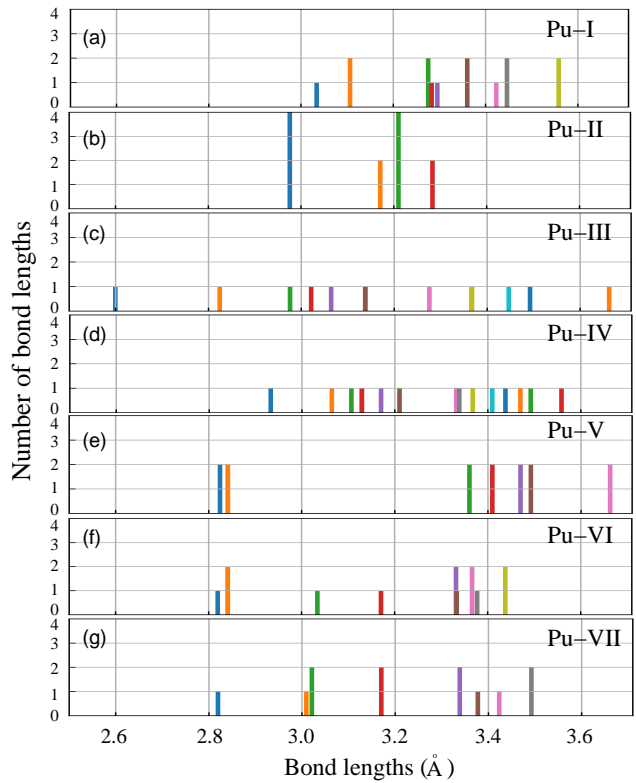}
 \caption{Distribution of bond lengths of each Pu-site in the $\beta$-Pu structure: (a) Pu-I, (b) Pu-II, (c) Pu-III, (d) Pu-IV, (e) Pu-V, (f) Pu-VI, and (g) Pu-VII.}
\label{fig:NN_dist}
\end{figure}

In Figs.~\ref{fig:NN_dist}(a)-(g) we illustrate the calculated bond length distribution, where only the Pu-Pu interatomic distances up to 3.71 \AA{} were taken into account (nearest-neighbor-bonds). Towards an approximate classification of bond lengths in $\beta$-Pu we define two groups of bonds, one group corresponding to the short bond lengths which takes into account bonds between 2.50 and 3.12 \AA{}, and the group of long bond lengths which includes bonds between 3.13 and 3.71 \AA{}. A similar classification of bond lengths was employed by Zachariasen and Ellinger in Ref.~\onlinecite{ellinger}. 
Within our classification we obtained the number of bonds listed in Table I.  In addition, we list in the same table the evaluated weighted average bond lengths among the groups for the distinct Pu-sites.

From table I one can notice that Pu-I site (Fig.~\ref{fig:NN_dist}(a)) has only three short bonds, with an average bond length of 3.08 \AA{}. The Pu-IV site (Fig.~\ref{fig:NN_dist}(d)) has a similar average bond length of 3.06 \AA{} and four short bonds. On the other hand, the Pu-III site (Fig.~\ref{fig:NN_dist}(c)) has the maximum number of short bonds, including even some bonds of 2.6 \AA{}. The average short bond length of this site is 2.90 \AA{}. Between these cases we have the intermediate sites, such as II, V, VI and VII, with four short bonds and \={d}$_{short} \approx 3$ \AA{}. 
%Further, the Pu-V (Fig.~\ref{fig:NN_dist}(e)) has four short bond lengths, with \={d}$_{short} = 2.83$ \AA{}. 
Similar to $\alpha$-Pu,~\cite{willsNatCom} these findings suggest that the Pu-5$f$ electrons in $\beta$-Pu will have multiple degrees of correlations due to the distinct local environments of each Pu-sites within the unit cell. We expect that the Pu-5$f$ electrons on sites III and V will have the largest hybridization with the other electronic states, and then a more itinerant character. In contrast, we expect the Pu-5$f$ states of Pu-I and Pu-IV sites to be more localized. 
\begin{table}[!htb]
\label{NN_res} 
\caption{Number of short and long bond lengths in $\beta$-Pu. \={d} denotes the weighted average bond lengths within each subgroup.}
\begin{ruledtabular}
\begin{tabular}{ccccc}
Pu site & short  & \={d}$_{short}$ (\AA{}) & long & \={d}$_{long}$(\AA{}) \\ \hline
$I$     &  3   & 3.08 &  11 & 3.39 \\
$II$    &  4   & 2.98 &  8  & 3.22 \\
$III$   &  5   & 2.90 & 7 & 3.39   \\
$IV$    &  4   & 3.06 & 10 & 3.38 \\
$V$     &  4   & 2.83 & 10 & 3.48  \\
$VI$    &  4    & 2.88  & 9 & 3.35 \\
$VII$   &  4   & 2.97 &  8 & 3.35 \\ \hline
\end{tabular}
\end{ruledtabular}
\end{table}

\begin{figure}[!htb]
\includegraphics[scale=0.4]{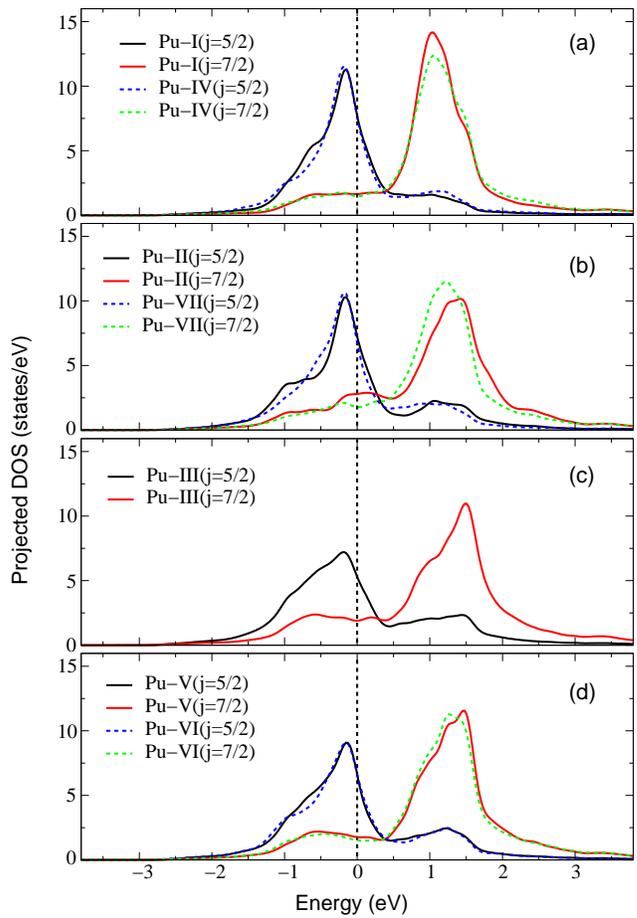}
 \caption{DFT(GGA) projected density of states on distinct Pu-sites. We show the projected density of states of sites I and IV (a), II and VII (b), III (c), and V and VI (d.) The projection on $5f_{5/2}$ and $5f_{7/2}$ states are denoted by $j = 5/2 $ and $j= 7/2$, respectively.}
\label{fig:pdosdft}
\end{figure}

Some effects of the distinct local environments of the Pu-sites can be observed in the electronic structure at DFT level.
In Fig.~\ref{fig:pdosdft} we show the calculated DFT(GGA) projected density of states for each Pu site. As a result of the strong spin-orbit coupling the Pu-5$f$ states splits into the Pu-5$f_{5/2}$ (total angular momentum $j = 5/2$) and the Pu-5$f_{7/2}$ (total angular momentum $j = 7/2$) manifolds. The position of the Pu-5$f_{5/2}$ states for each inequivalent Pu-site is roughly the same, i.e. they appear just below the Fermi energy. However, the peaks associated with the Pu-5$f_{7/2}$ states vary according to their local environment. We observe the Pu-5$f_{7/2}$ states of Pu-I sites at approximately 1 eV, whereas the same states of Pu-III sites are peaked at around 1.5 eV. Although these states are mainly located between 0.5 and 2 eV, it is important to mention that they have a finite, though small, contribution to the electronic states at the Fermi energy.

\begin{figure}[!htb]
 \includegraphics[scale=0.38]{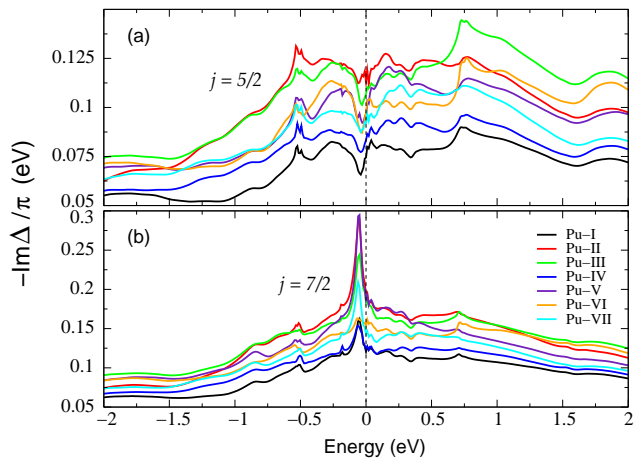}
 \caption{DFT(GGA) hybridization function of $5f$ states associated with each distinct Pu-sites. We show -$\frac{Im \Delta}{\pi}$ of (a) 5$f_{5/2}$ and (b) 5$f_{7/2}$.}
 \label{fig:hybrid_dft}
 \end{figure}

From the electronic structure point of view, some of the Pu-sites can be roughly divided into different groups. Within each group the Pu-5$f_{5/2}$ and Pu-5$f_{7/2}$ density of states exhibit similar positions and intensities. As can be noticed in Figs.~\ref{fig:pdosdft}(a) and (d) these states exhibit very similar features among sites I and IV, and V and VI, respectively. In the case of Pu-II the 5$f_{5/2}$ states are similar to those of Pu-VII site whereas the 5$f$ electrons in Pu-III site exhibit a distinct feature.
These findings can be explained in part due to the bond length distribution of each site and the corresponding hybridization of these sites with the remaining electrons. As listed in table~I, these sites have essentially the same average short bond lengths, i.e. for sites I and IV \={d}$_{short} \sim 3.1$ \AA{}, for sites V and VI \={d}$_{short} \sim 2.9$ \AA{}, and for sites II and VII  \={d}$_{short} \sim 3$ \AA{}. In fact, the DFT hybridization function shown in Fig.~\ref{fig:hybrid_dft} captures the effects of the different local environments of each Pu-site. From Figs.~\ref{fig:hybrid_dft}(a) and (b) we observe that the 5$f$ electrons in Pu-I and IV sites are the ones where both Pu-5$f_{5/2}$ and 5$f_{7/2}$ states are less hybridized, and then more localized. In contrast, the Pu-5$f$ states in sites II, III and V are the ones which are more hybridized and itinerant.

\begin{figure}[!htb]
 \includegraphics[scale=0.4]{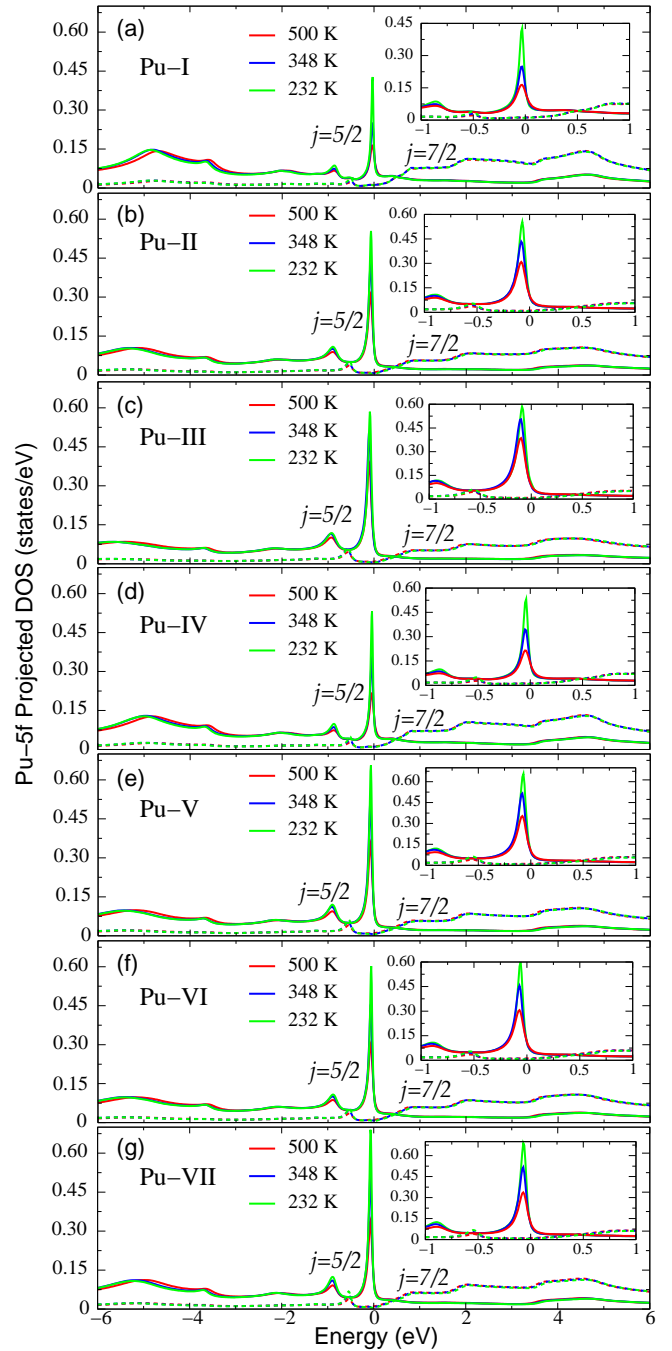}
 \caption{DFT+DMFT(OCA) obtained Pu-5$f$ projected density of states at 500 K (red), 348 K (blue), and 232 K (green), of Pu-I (a), Pu-II (b), Pu-III (c), Pu-IV (d), Pu-V (e), Pu-VI (f), and Pu-VII (g) sites. Continuous (dashed) lines denote the Pu-5$f_{5/2}$ (Pu-5$f_{7/2}$) projected density of states. Insets display the projected density of states in a smaller energy window (E$_{F}\pm 1$ eV).}
 \label{fig:pdosdmft_oca}
\end{figure}

We now turn to the investigation of the effects of electronic correlations on the electronic structure of $\beta$-Pu. We performed DFT+DMFT(OCA) calculations for three different temperatures, namely, 500 K, 348 K and 232 K. In Fig.~\ref{fig:pdosdmft_oca} we show the DFT+DMFT Pu-5$f$ projected density of states for all temperatures for all the seven Pu-sites.
When compared with the DFT based projected density of states, we observe that both Pu-5$f_{5/2}$ and Pu-5$f_{7/2}$ states are strongly affected by the electronic correlations.
The calculated Pu-5$f$ average occupancy of $\langle n_{f} \rangle = 5.26$, indicates that Pu-5$f$ electrons are in a mixed valence state in $\beta$-Pu, similar to the $\delta$ phase.
The electronic correlations taken into account within DMFT give rise to strongly renormalized quasiparticle peaks near the Fermi energy which appears due to antiferromagnetic interaction of the Pu-5$f$ electrons with the surrounding conduction electrons. These heavy quasiparticles states, which are of Pu-5$f_{5/2}$ character, exhibit a strong temperature dependence where their intensity increases upon decrease in temperature (see insets in Fig.~\ref{fig:pdosdmft_oca}). This behavior was also reported in previous studies on the family of heavy fermion compounds Ce-115~\cite{shimScience} and on PuCoGa$_5$.~\cite{whbpu115}
Another hallmark of the many-body effects in this system is the appearance of Hubbard bands at higher energies and quasiparticle multiplets in the Pu-5$f$ spectral function. 
The position of the lower Hubbard bands are weakly site-dependent, where they appear at around -5 eV for the spectral functions associated with sites Pu-I and Pu-IV, whereas for sites Pu-III they are centered at -5.7 eV. The position of the quasiparticle multiplets, in contrast, are site independent. They can be notice at -0.9 eV (Pu-5$f_{5/2}$ character) and at -0.6 eV (Pu-5$f_{7/2}$ character) in the Pu-5$f$ spectral function. It is noteworthy that similar multiplets have been reported in previous studies employing DFT+DMFT within OCA  for the $\delta$ phase of elemental Pu,~\cite{shimOCAPu} and in Pu-based compounds.~\cite{chuckpu,whbpu115}

\begin{figure}[!htb]
 \includegraphics[scale=0.40]{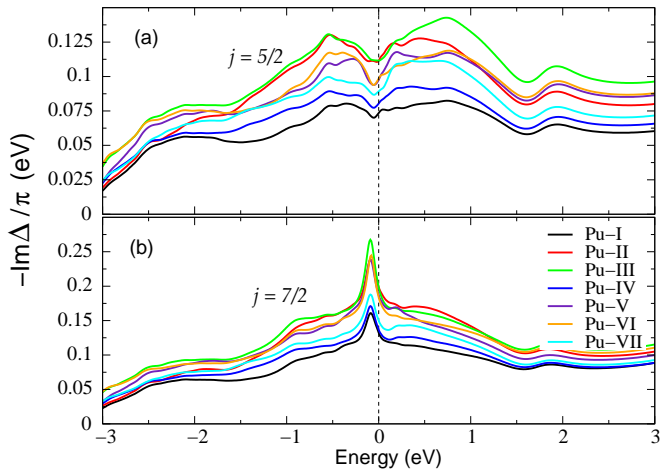}
 \caption{DFT+DMFT hybridization function, at 500 K, of Pu-5$f$ states at all the Pu-sites. We show -Im$\Delta$/$\pi$ for (a) Pu-5$f_{5/2}$(a) and (b) Pu-5$f_{7/2}$ states.}
 \label{fig:hybrid_dmft}
\end{figure}

Although the position of the Hubbard bands and the quasiparticle multiplets does not shown a clear site-dependence, the Pu-$5f$ spectral function around the Fermi energy exhibits some site selectivity. As can be seen in Fig.~\ref{fig:pdosdmft_oca}, the Pu-$5f_{5/2}$ quasiparticle peaks at Pu-I and Pu-III sites have very different intensities. Overall, we find more intense quasiparticle peaks for the Pu sites where the 5$f$ electrons are more hybridized with the conduction electrons, such as at Pu-II, Pu-III, Pu-V, and Pu-VII sites. In fact, both DFT and DFT+DMFT hybridization functions (Fig.~\ref{fig:hybrid_dmft}), demonstrate that Pu-5$f_{5/2}$ states at these sites are more hybridized than the Pu-5$f$ electrons at Pu-I site.   

These site selective many-body effects reflect the different Kondo scales of Pu-5$f$ electrons within $\beta$-Pu. The Kondo temperature, which is the energy scale associated to the buildup of coherence, depends  exponentially on the intensity of the hybridization function.~\cite{hewsonbook} Therefore, those sites at which the Pu-5$f_{5/2}$ states are more hybridized (larger intensity of hybridization function) will start to buildup coherence at higher temperatures than the sites where the Pu-5$f$ states have a weak hybridization.
It is interesting to note that the DFT hybridization function can be used  as a simple quantity to determine if there are multiple Kondo temperatures within highly anisotropic crystal structures. As can be notice in Fig.~\ref{fig:hybrid_dmft}, the correlations enhance the distinction of the hybridization among the Pu-sites, and is also a good indicate of site selectivity.

The multiple degrees of correlations in $\beta$-Pu can be also noticed in the quasiparticle weights of Pu-5$f_{5/2}$ electrons (see upper part of Table~II). 
Although the one crossing approximation is known to overestimate the many-body renormalizations in 5$f$ systems,~\cite{zhu} one can compare the quasiparticle weights associated to Pu-5$f$ states at the Pu-sites. For instance, the Z$_{5/2}$ of Pu-5f$_{5/2}$ electrons at Pu-III  is twice as large as the Z$_{5/2}$ of Pu-I and Pu-IV sites. This indicates that in $\beta$-Pu the Pu-5$f_{5/2}$ electrons are more correlated at sites I and IV, and are more itinerant at Pu-III which in turn is more hybridized. Overall, our DFT+DMFT(OCA) calculations indicate the existence of site selective correlations in $\beta$-Pu where the Pu-5$f$ electrons have multiple degrees of correlations, which adds another system to the group of materials which exhibits this kind of phenomena.~\cite{parkPRL,willsNatCom}

\begin{table}[!htb]
\label{Zns_res} 
\caption{Quasiparticle weights and occupancies of Pu-5$f_{5/2}$ and Pu-5$f_{7/2}$ states in $\beta$-Pu at 500 K.}
\begin{ruledtabular}
\begin{tabular}{ccccc}
Pu site & $Z_{5/2}$ & $Z_{7/2}$ & $n_{5/2}$ & $n_{7/2}$ \\ \hline
        &           &  $J_{H} = 0.512$ eV &    &   \\  
$I$     &  0.05    &  0.34 &  4.05 & 1.08\\
$II$    &  0.08    &  0.17 &  4.07 & 1.15 \\
$III$   &  0.10    &  0.13 & 4.11 & 1.17 \\
$IV$    &  0.05    &  0.28  & 4.05 & 1.10    \\
$V$     &  0.08    &  0.17  & 4.09 & 1.15 \\
$VI$    &  0.07    &  0.20  & 4.07 & 1.15 \\
$VII$   &  0.07    &  0.20 & 4.08 & 1.12 \\ \hline 
        &           &  $J_{H} = 0$ &    &   \\
$I$     &  0.10    &  0.59 &  4.97 & 0.16 \\
$II$    &  0.17    &  0.30 &  5.03 & 0.18 \\
$III$   &  0.21    &  0.44 &  5.07 & 0.19 \\
$IV$    &  0.12    &  0.54 &  4.99 & 0.17 \\
$V$     &  0.17    &  0.38 &  5.03 & 0.19 \\
$VI$    &  0.15    &  0.50 &  5.02 & 0.19 \\
$VII$   &  0.18    &  0.46 &  5.04 & 0.17 \\
\hline
\end{tabular}
\end{ruledtabular}
\end{table}

\subsection{Origin of the orbital differentiation}

Resolving by orbitals the  calculated Pu-5$f$ spectral functions  reveals  marked  differences between the spectra of the Pu-5$f_{5/2}$ and Pu-5$f_{7/2}$ states. The former gives rise to a strongly renormalized quasiparticle state near the Fermi energy, a quasiparticle multiplet peak at -0.9 eV and to the lower Hubbard bands at higher energies.  The Pu-$5f_{7/2}$ states, in contrast, are  strongly suppressed and gives rise to upper Hubbard bands, besides the quasiparticle multiplet around -0.6 eV. The Pu-5$f_{7/2}$ spectral function is essentially gapped around the Fermi energy and also temperature independent, which rules out any sign of buildup of coherence down to 232 K. These findings indicate a drastic difference of Kondo energy scales, for all the Pu sites, of the Pu-5$f_{5/2}$ and Pu-5$f_{7/2}$ states, where we expect T$_{K,7/2} <<$ T$_{K,5/2}$. It is important to mention that the Pu-5$f_{5/2}$ states are occupied by about four electrons while the Pu-5$f_{7/2}$ have one electron, and the occupancies are almost temperature independent.

\begin{figure}[!htb]
 \includegraphics[scale=0.4]{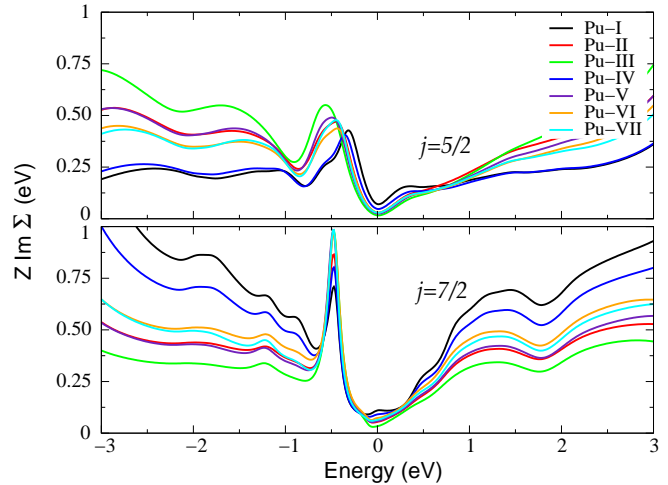}
 \caption{Renormalized imaginary part of self-energies of (a) Pu-5$f_{5/2}$ and (b) Pu-5$f_{7/2}$ states at 500 K for each Pu-site.}
 \label{fig:scatt_rate}
\end{figure}

This can be understood in terms of the  larger hybridization of the Pu-$5f_{7/2}$ states, in the low energy region (Fig.~\ref{fig:hybrid_dmft}(b)), which  leads to larger quasiparticle weights than the ones associated to Pu-5$f_{5/2}$ states. For instance, $Z_{7/2}/Z_{5/2} \approx 7$ at Pu-I site while $Z_{7/2}/Z_{5/2} \approx 3$ at site Pu-VII. 
To understand why these states have so different spectral functions we now turn to the imaginary parts of the DMFT self-energies for both orbitals.
In Fig.~\ref{fig:scatt_rate} we display the $Z$Im$\Sigma(\omega)$ for each Pu-5$f$ state.  Both Pu-5$f_{5/2}$ and 5$f_{7/2}$ give rise to small scattering rates near zero frequency, i.e. $Z$Im$\Sigma(0)$  are small, in the vicinity of the Fermi energy, where the quasiparticle peak  of  Pu-$5f_{5/2}$ appears. However, there is a strong enhancement  of the imaginary part of the Pu-5$f_{7/2}$ self-energy at higher energies  (near the energies of the Pu-$5f_{7/2}$ bands), which leads to a strong suppression the Pu-5$f_{7/2}$ spectral function and explains its incoherent nature. Therefore, the orbital differentiation of the Pu-5$f$ can be attributed due to the difference between the renormalized imaginary parts of Pu-5$f_{5/2}$ and Pu-5$f_{7/2}$ self-energies, where the latter is strongly enhanced at higher energies. 

\subsection{Effects of Hund's coupling}

The presence of the atomic multiplet structure in  the one particle spectral function of the actinides has been stressed in earlier publications. These features, were denoted quasiparticle multiplets in Ref.~\onlinecite{chuckpu}, and this unique feature of the actinides deserves the name Racah materials a term introduced in Ref.~\onlinecite{shick_racah}. 

Here we investigate this effect further, by turning on and off  the effects of the Hund's coupling ($J_{H}$) as was done for the iron pnictides in Ref.~\onlinecite{Haule_Hunds}. Hence we performed additional calculations of the  spectra of Pu-5$f$ electrons in $\beta$-Pu with  $J_{H}=0$ at 500 K. The evaluated quasiparticle weights and corresponding orbital occupancies are shown in the lower part of Table II. 

The quasiparticle weights of the Pu-5$f_{5/2}$ states,  obtained  with $J_{H}=0$, are approximately twice as larger as the quasiparticles weights obtained with finite value of $J_{H} = 0.512$ eV, i.e. $Z_{5/2}(J_{H}= 0) \approx 2 Z_{5/2}(J_{H}=0.512)$. This indicates that a considerable amount of correlations on the Pu-5$f_{5/2}$ states comes from the Hund's coupling. It is important to mention that the Pu-5$f$ electrons, in absence of the Hund's coupling, are still quite correlated as indicated by the obtained smaller values of $Z$ with $J_{H} = 0$. This means that the Coulomb interaction among the Pu-5$f$ electrons introduce a considerable amount of correlations into the system. 
The orbital occupancies are strongly modified. As can be seen in Table~II the Pu-$5f_{5/2}$ states gain one electron while the Pu-$5f_{7/2}$ states become essentially empty. 

\begin{figure}[!htb]
 \includegraphics[scale=0.42]{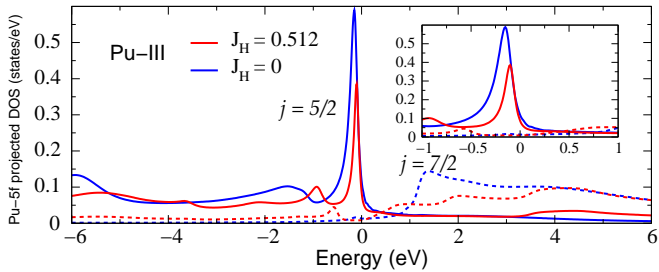}
 \caption{DFT+DMFT(OCA) based Pu-$5f$ projected density of states of Pu-III site at 500 K. These results were obtained for $J_{H} = 0.512$ eV (red) and $J_{H} = 0$ (blue). Continuous (dashed) lines denote the Pu-5$f_{5/2}$ (Pu-5$f_{7/2}$) projected density of states. In the inset we show the projected density of states in a smaller energy window (E$_{F}\pm 1$ eV).}
 \label{fig:pdoscomp}
\end{figure}

The Hund's coupling has also a strong influence on the Pu-5$f$ spectral function, as can be seen in Fig.~\ref{fig:pdoscomp}. When $J_{H}=0$, the Pu-5$f_{5/2}$ quasiparticle peak becomes more intense and a little bit broader, which suggests that the Hund's coupling reduces the Kondo temperature of these states. Another drastic difference is that the quasiparticle multiplets disappear when the Hund's coupling is neglect. As pointed out in previous calculations by Shick et al.,~\cite{shick_racah,Lichtenstein_Racah} similar peaks appear in the spectra of $\delta$-Pu and Pu-based compounds due to $5f^{6}\rightarrow 5f^{5}$ transitions, and are analogs to the Racah peaks associated with transitions between Racah multiplets.~\cite{racah}  

Therefore, our results demonstrate that $\beta$-Pu can be considered a realization of a Hund's-Racah metal, and that the Hund's coupling is essential to describe the electronic excitations of this type of correlated metallic systems.

\section{Conclusions}

In summary, we  have investigated the electronic structure of $\beta$-Pu, which has a monoclinic structure with seven distinct crystallographic sites. Our calculations reveal very interesting physics, rather than a weakly correlated phase, we find within DFT+DMFT using OCA as an impurity solver, large mass renormalizations. There is less site selectivity than in  $\alpha$-Pu, with site I, the most correlated one.  More important, our calculations suggest that the Pu-5$f_{5/2}$ and Pu-5$f_{7/2}$ states have very distinct spectral functions. We notice this effect for PuCoGa$_5$ and for UO$_2$ and its natural to conjecture that this effect is generic feature in actinide materials. We also find that Hund's coupling is essential to the appearance of the quasiparticle multiplets in the Pu-5$f$ spectral function and for determining the level of correlations in $\beta$-Pu, adding additional realizations of Hund's-Racah metallic systems.

\section{Acknowledgments}
This work  was supported by U.S. DOE BES under Grant No. DE-FG02-99ER45761.


\begin{thebibliography}{100}

%% _review

\bibitem{moore_review} K. T. Moore and G. van der Laan,
Rev. Mod. Phys. {\bf 81}, 235 (2009).


%%% specific heat 
\bibitem{lashleyPRL} J. C. Lashley, J. Singleton, A. Migliori, J. B. Betts, R. A. Fisher, J. L. Smith, and R. J. McQueeney,
Phys. Rev. Lett. {\bf 91}, 205901  (2003).

\bibitem{havelaPRB} L. Havela, P. Javorsk\'y, A. B. Shick, J. Koloren\v{c}, E. Colineau, J. Rebizant, F. Wastin, J.-C. Griveau, L. Jolly, G. Texier, F. Delaunay, and N. Baclet,
Phys. Rev. B {\bf 82}, 155140 (2010).



%%% magn. susc. delta-Pu
\bibitem{lashley} J. C. Lashley, A. Lawson, R. J. McQueeney, and G. H. Lander,
Phys. Rev. B {\bf 72}, 054416 (2005).

%%% magn. alpha delta Pu
\bibitem{heffnerPRB} R. H. Heffner, G. D. Morris, M. J. Fluss, B. Chung, S. McCall, D. E. MacLaughlin, L. Shu, K. Ohishi, E. D. Bauer, J. L. Sarrao, W. Higemoto, and T. U. Ito,
Phys. Rev. B {\bf 73}, 094453 (2006).


%%%% prev. DFT calc.
\bibitem{dftpu1} P. S\"{o}derlind, O. Eriksson, B. Johansson, and J. M. Wills,
Phys. Rev. B {\bf 50}, 7291 (1994).

\bibitem{dftpu2} I. V. Solovyev, A. I. Liechtenstein, V. A. Gubanov, V. P. Antropov, and O. K. Andersen,
Phys. Rev. B {\bf 43}, 14414 (1991).

%%% previous DMFT on elemental Pu
\bibitem{elihu} S. Y. Savrasov, G. Kotliar, and E. Abrahams,
Nature {\bf 410}, 793 (2001).

\bibitem{xidai_Science} X. Dai, S. Y. Savrasov, G. Kotliar, A. Migliori, H. Ledbetter, and E. Abrahams,
Science {\bf 300}, 953 (2003).

\bibitem{marianetti_PRL} C. A. Marianetti, K. Haule, G. Kotliar, and M. J. Fluss,
Phys. Rev. Lett. {\bf 101}, 056403 (2008). 

\bibitem{amadon_PRB} B. Amadon,
Phys. Rev. B {\bf 94}, 115148 (2016).

\bibitem{shimOCAPu} J. H. Shim, K. Haule, and G. Kotliar,
Nature {\bf 446}, 513 (2007).

%% U, J - delta Pu
\bibitem{janoschek} M. Janoschek, P. Das, B. Chakrabarti, D. L. Abernathy, M. D. Lumsden, J. M. Lawrence, J. D. Thompson, G. H. Lander, J. N. Mitchell, S. Richmond, M. Ramos, F. Trouw, J.-X. Zhu, K. Haule, G. Kotliar, and E. D. Bauer,
Sci. Adv. {\bf 1} (2015).

%% alpha Pu
\bibitem{willsNatCom} J.-X. Zhu, R. C. Albers, K. Haule, G. Kotliar, and J. M. Wills,
Nat. Commun.{\bf 4}, 2644 (2013). 


%% Peierls distortion DFT 
\bibitem{soderling_PRBPeierls} P. S\"{o}derlind, J. M. Wills, B. Johansson, and O. Eriksson,
Phys. Rev. B {\bf 55}, 1997 (1997). 

%% review soderlind Gabi

\bibitem{review_soderlind} P. S\"{o}derlind and G. Kotliar, 
\emph{Theoretical Electronic Structure of Plutonium Metal and Alloys.} In Plutonium Handbook, Second Edition: D. L. Clark, D. A. Geeson, Jr. R. Hanrahan, Eds.; American Nuclear Society: Chicago, IL submitted for publication 2018; Chapter 14.

%%%


\bibitem{lanataPRX} N. Lanat\`a, Y. Yao, C.-Z. Wang, K.-M. Ho, and G. Kotliar,
Phys. Rev. X {\bf 5}, 011008 (2015).

\bibitem{soderlindPRL} P. S\"{o}̈derlind and B. Sadigh,
Phys. Rev. Lett. {\bf 92}, 185702 (2004).

\bibitem{wallacePRB} D. C. Wallace,
Phys. Rev. B {\bf 58}, 15433 (1998).


\bibitem{suzukiPRB} Y. Suzuki, V. R. Fanelli, J. B. Betts, F. J. Freibert, C. H. Mielke, J. N. Mitchell, M. Ramos, T. A. Saleh, and A. Migliori,
Phys. Rev. B {\bf 84}, 064105 (2011).

%% pu19Os

\bibitem{havela_puos} L. Havela, S. Ma\v{s}kov\'a, J. Koloren\v{c}, E. Colineau, J. C. Griveau, and R. Eloirdi,
J. Phys.: Condens. Matter {\bf 30}, 085601 (2018).


%%% orb selec.
\bibitem{anisimov_ruth} V. I. Anisimov, I. A. Nekrasov, D. E. Kondakov, T. M. Rice, and M. Sigrist,
Eur. Phys. J. B {\bf 25}, 191 (2002).


\bibitem{ziping} Z. P. Yin, K. Haule, and G. Kotliar,
Nat. Materials {\bf 10}, 932 (2011).

\bibitem{lanata_iron} N. Lanat\`a, H. U. R. Strand, G. Giovannetti, B. Hellsing, L. de' Medici, and M. Capone,
Phys. Rev. B {\bf 87}, 045122 (2013).

\bibitem{demedici_PRL} L. de’ Medici, G. Giovannetti, and M. Capone,
Phys. Rev. Lett. {\bf 112}, 177001 (2014).

\bibitem{miao} H. Miao, Z. P. Yin, S. F. Wu, J. M. Li, J. Ma, B.-Q. Lv, X. P. Wang, T. Qian, P. Richard, L.-Y. Xing, X.-C. Wang, C. Q. Jin, K. Haule, G. Kotliar, and H. Ding,
Phys. Rev. B {\bf 94}, 201109 (2016).




%%% UO2
\bibitem{lanata_uo2} N. Lanat\`{a}, Y. Yao, X. Deng, V. Dobrosavljevi\'c, and G. Kotliar,
Phys. Rev. Lett. {\bf 118}, 126401 (2017).

%%% UO2 under pressure
\bibitem{huang_uo2} L. Huang, Y. Wang, and P. Werner,
EPL {\bf 119}, 57007 (2017).

%%% Cf
\bibitem{huang_cf} L. Huang and H. Lu,
arXiv:1807.07505 (2018).

\bibitem{whbpu115} W. H. Brito, S. Choi, Y. X. Yao, and G. Kotliar,
Phys. Rev. B {\bf 98}, 035143 (2018).


%%% racah metals
\bibitem{shick_racah} A. B. Shick, J. Kolorenc, J. Rusz, P. M. Oppeneer, A. I. Lichtenstein, M. I. Katsnelson, and R. Caciuffo,
Phys. Rev. B {\bf 87}, 020505 (2013).


\bibitem{Lichtenstein_Racah} A. B. Shick, L. Havela, A. I. Lichtenstein, and M. I. Katsnelson,
Sci. Rep. {\bf 5}, 15429 (2015).

\bibitem{hauleWK} K. Haule, C.-H. Yee, and K. Kim,
Phys. Rev. B {\bf 81}, 195107 (2010).

\bibitem{haulepage} \url{http://hauleweb.rutgers.edu/tutorials}

% pbe
\bibitem{pbe} J. P. Perdew, K. Burke, and M. Ernzerhof,
Phys. Rev. Lett. {\bf 77}, 3865 (1996).

%wien2k
\bibitem{wien} P. Blaha, K. Schwarz, G. K. H. Madsen, D. Kvasnicka, and J. Luitz,
\emph{WIEN2K, An Augmented Plane Wave + Local Orbitals Program for Calculating Crystal Properties} (Karlheinz Schwarz, Techn. Universit\"{a}t Wien, Austria, 2001).

%oca 
\bibitem{pruschke} Th. Pruschke and N. Grewe,
Z. Phys. B {\bf 74}, 439 (1989).

%% double count.
\bibitem{anisimovEdc} V. I. Anisimov, F. Aryasetiawan, and A. I. Lichtenstein,
J. Phys.: Condens. Matter {\bf 9}, 767 (1997).


%% experimental crystal structure of beta-Pu
\bibitem{ellinger} W. H. Zachariasen and F. H. Ellinger,
Acta Cryst. {\bf 16}, 369 (1963).

%% oca Ce-115
\bibitem{shimScience} J. H. Shim, K. Haule, and G. Kotliar,
Science {\bf 318}, 1615 (2007).


%% Pu chalcogenides
\bibitem{chuckpu} C.-H. Yee, G. Kotliar, and K. Haule,
Phys. Rev. B {\bf 81} 035105 (2010).

%% hewson book
\bibitem{hewsonbook} A. C. Hewson, 
\emph{The Kondo problem to heavy fermions (Cambridge University Press, 1993).}

%% pu115 oca
\bibitem{zhu} J.-X. Zhu, P. H. Tobash, E. D. Bauer, F. Ronning, B. L. Scott, K. Haule, G. Kotliar, R. C. Albers, and J. M. Wills,
Europhys. Lett. {\bf 97}, 57001 (2012).

%% nickelates
\bibitem{parkPRL} H. Park, A. J. Millis, and C. A. Marianetti,
Phys. Rev. Lett. {\bf 109}, 156402 (2012).


\bibitem{Haule_Hunds} K. Haule and G. Kotliar,
New J. Phys. {\bf 11}, 025021 (2009).

\bibitem{racah} G. Racah,
Phys. Rev. {\bf 76}, 1352 (1949).


\end{thebibliography}
\end{document}